\documentclass[11pt]{article}
\usepackage[margin=1in]{geometry}
\usepackage[utf8]{inputenc}
\usepackage{microtype}
\usepackage{amsmath,amssymb,amsthm,mathtools}
\usepackage{booktabs,enumitem}
\usepackage{graphicx}
\usepackage{tikz,url}
\usetikzlibrary{arrows.meta,positioning,calc,fit}
\usepackage[hidelinks]{hyperref}
\usepackage{cleveref}

\newtheorem{theorem}{Theorem}[section]
\newtheorem{lemma}[theorem]{Lemma}

\newcommand{\rank}{\operatorname{rank}}
\newcommand{\eps}{\varepsilon}
\newcommand{\one}{\mathbf 1}

\title{The $k$-Clique hypothesis is dead, for good}
\author{
Virginia Vassilevska Williams\\\emph{MIT}
\and
Or Zamir\\\emph{Tel Aviv University}}
\date{}

\begin{document}
\maketitle

\begin{abstract}
We present an algorithm that solves the $k$-Clique problem for any large enough~$k$ and any $n$-vertex graph in $n^{\left(\frac{\omega}{3}-c\right)k}$ time, where $\omega\geq 2$ is the matrix multiplication exponent and $c>0$ is a positive constant \emph{independent} of both $\omega$ and $k$.
This refutes the long-standing $k$-Clique conjecture, regardless of the value of the matrix multiplication constant, even if $\omega=2$.
In fact, our algorithm solves the more general problem of counting weighted $k$-Cliques.
\end{abstract}

\section{Introduction}
Finding cliques in graphs is a classic problem in computer science with an abundance of applications. The problem is NP-hard when the desired clique size $k$ is part of the input. When $k$ is a fixed small constant, there is a well-known way to beat the brute-force $k$-tuple enumeration algorithm. More than 40 years ago, Ne\v{s}etril and Poljak~\cite{NesetrilPoljak1985} showed how to reduce $k$-clique in $n$ node graphs to triangle (i.e. $3$-clique) detection in $O(n^{k/3})$ node graphs, when $k$ is divisible by $3$. This leads to 
 an $n^{\omega k/3+o(1)}$-time algorithm for $k$ divisible by three~\cite{NesetrilPoljak1985}, and only adds at most $2/3$ to the exponent\footnote{If $k=t \mod 3$, we can reduce the problem to $n^t$ instances of $k-t$ clique, and the overall exponent for $k$-clique becomes $\omega(k-t)/3+t=\omega k/3 +(3-\omega)t/3\leq \omega k/3 +t/3\leq \omega k/3+2/3$.} for other $k$. Here $\omega\leq 9/4$ denotes the exponent of square matrix multiplication (see e.g. \cite{openai26,abz,shp,strassen,cw90,moreasymmetry}).
 
 The exponent for $k$-clique detection has remained unchallenged until today, except for small improvements using rectangular matrix multiplication for the cases when $k$ is not divisible by $3$ (e.g. \cite{EisenbrandG04}). This motivated the somewhat dubious, yet perfectly consistent with the state of affairs, $k$-Clique Hypothesis~\cite{Woeginger04,AbboudBW18}:
 \begin{center}
The {\em $k$-Clique Hypothesis}: in the word-RAM model with $O(\log n)$ bit words, for $k$ divisible by $3$, $k$-clique detection in $n$ node graphs requires $n^{\omega k/3-o(1)}$ time.
 \end{center}
Stronger versions~\cite{DalirrooyfardVW21} even conjectured the optimality of the rectangular matrix multiplication solution for values of~$k$ not divisible by three.
%(There are stronger versions too~\cite{DalirrooyfardVW21}: that the exponent for any constant $k\geq 3$, the exponent is, within $o(1)$, exactly $\omega(\lfloor k/3\rfloor,\lceil k/3\rceil,k-\lfloor k/3\rfloor-\lceil k/3\rceil)$ where $\omega(a,b,c)$ is the exponent of $n^a\times n^b$ by $n^b\times n^c$ matrix multiplication. This extends the hypothesis for $k$ not divisible by $3$.) 

Fine-grained complexity then developed many reductions from $k$-Clique to other algorithmic problems (e.g. \cite{AbboudBW18,Chang16,BringmannW17,JayaramS17,AbboudBBK17,NogamiNT26,LincolnWW18,DalirrooyfardVW21,DalirrooyfardW22,Chan10,FischerKRS25,FritschKRS26,GorbachevK23}), showing that in order to make further progress on these problems, one must first refute the $k$-Clique Hypothesis. 

The $k$-Clique Hypothesis is usually used for problems whose best running times rely on matrix multiplication (MM) in some way. To show that MM is likely necessary for the best algorithm for a problem $\pi$, one  reduces to $\pi$, either MM itself, or another problem that is believed to require MM. Such problems include Boolean Matrix Multiplication (BMM, whose best algorithms are the same as for MM), triangle detection (which is subcubically equivalent to BMM~\cite{WilliamsW10,WilliamsW18}) or more generally $k$-Clique either for some other fixed $k$ (such as $k=4$) or for any constant $k\geq 3$, relying on the $k$-Clique Hypothesis.
The latter route is particularly useful when the reduction has some polynomial overhead which the high conjectured running time of $k$-Clique lets you overcome. 

For instance, in Abboud, Backurs and Vassilevska W.'s reduction from $k$-Clique to Context Free Grammar (CFG) Recognition~\cite{AbboudBW18}, an $n$-node graph is encoded as a string of length $O(n^{k/3+1})$ for a single fixed grammar, a factor of $n$ more than the $n^{k/3}$ that a tight reduction would produce. An $O(n^{\omega-\eps})$-time CFG recognizer would then solve $k$-Clique in $O(n^{(\omega-\eps)(k/3+1)})=O(n^{\omega k/3-\eps k/3+\omega-\eps})$ time. This is faster than $n^{\omega k/3}$ only when $k>3(\omega-\eps)/\eps$ and, under the $k$-Clique Hypothesis that makes an assertion for all $k$, Valiant's $O(n^\omega)$-time parser is optimal. If one were trying to get the reduction to work from small $k$, such as from triangle, it would not work because of the extra $+1$ overhead in the reduction exponent.
%, but the $k$-Clique Hypothesis allows one to use any large enough $k$.
There are many such examples, where a reduction from a fixed size clique such as triangle (or from BMM) would not quite work due to the overhead, so the $k$-Clique Hypothesis became useful.

In 2004, Woeginger~\cite{Woeginger04} challenged algorithms researchers to beat the Ne\v{s}etril-Poljak running time for some fixed $k$. Over two decades passed, the running time remained untouched, the hypothesis gained more traction and more fine-grained problems were found to be $k$-clique hard.

In this paper we strongly refute this hypothesis, and we bring forth hope that the $k$-clique hard problems can also have faster algorithms. We prove:

\begin{theorem}\label{thm:kclique}
There exists constants $K_0$ and $\delta>0$, independent of the matrix multiplication exponent~$\omega\geq 2$, such that for every $k\geq K_0$, $k$-Clique can be solved in $O(n^{(\omega-\delta)k/3})$ time. In fact, the algorithm can return the number of $k$-cliques in the given graph.
\end{theorem}

We did not try to optimize $K_0$ or $\delta$, as we eventually discuss in Section~\ref{sec:discuss}.

Recent breakthroughs have refuted several central conjectures in fine-grained complexity, including the APSP and 3SUM hypotheses~\cite{AlmanVW2026}, the Exponential Time Hypothesis~\cite{OpenAIETH2026}, and the Hyperclique Hypothesis~\cite{Fischer2026Hyperclique,AlmanLiPantPrattWilliams2026,ChanItakuraMaYu2026}. The $k$-Clique Hypothesis presents an additional difficulty: unlike these conjectures, its predicted running time depends on $\omega$, whose exact value is unknown. Thus, even an improvement over the best known $k$-Clique algorithm could cease to contradict the hypothesis if a subsequent improvement in matrix multiplication lowered the known upper bound on $\omega$. Our result avoids this moving target: the exponent saving is relative to the true value of $\omega$, and therefore remains a refutation of the hypothesis regardless of any future progress on matrix multiplication, including in the case that it would be proven that~$\omega=2$.

In order to obtain improved $k$-Clique running times, we solve a potentially harder {\em weighted $k$-clique sum} problem that generalizes counting the number of $k$-cliques: Every edge $e$ of the host graph $G$ has a weight $w(e)$, the weight of a $k$-clique in $G$ is the product of its edge weights, and we want to compute the sum of all weights of $k$-cliques in $G$.
The Ne\v{s}etril-Poljak style reduction works just as well for weighted $k$-clique sum, and the problem in $n$ node graphs can be reduced  to weighted $\ell$-clique sum in $O(n^{k/\ell})$ node graphs (when $k$ is divisible by $\ell$). Thus, to get a faster weighted $k$-clique sum algorithm for all large enough $k$, it suffices to get a faster weighted $\ell$-clique sum algorithm for some constant $\ell$. We do this for $\ell=9$.

In fact, we give an improved algorithm that computes a weighted sum over the copies of any given $9$-node pattern $H$ in an $n$-node graph $G$ (see Section~\ref{sec:algo} for a definition for this more general problem):

\begin{theorem}\label{thm:main}
There is an absolute constant $\eps>0$ such that, over any field of characteristic zero, 
for any $n$-node host graph $G$ and every pattern graph $H$ on nine vertices
the weighted sum over the copies of $H$ in $G$ can be computed using $n^{3\omega-\eps+o(1)}$ arithmetic operations.
\end{theorem}

%\begin{corollary}\label{cor:clique}
%There is an absolute constant $c>0$ such that, for every fixed positive integer $s$, weighted $9s$-clique counting can be performed using
%\[
% n^{(\omega/3-c)9s+o(1)}
%\]
%arithmetic operations. The constant $c$ is independent of $s$ and of $\omega$.
%\end{corollary}

As $k$-clique in unweighted graphs can be converted to the corresponding weighted sum problem with small weights, the arithmetic complexity implies a similarly efficient algorithm in the word-RAM model, and combining the algorithm with the Ne\v{s}etril-Poljak reduction, we obtain Theorem~\ref{thm:kclique} as a corollary.

Our proof makes central use of an algebraic decomposition lemma that we distilled, with a publicly available AI model's assistance, out of the recent ETH refutation of OpenAI~\cite{OpenAIETH2026}.
We separate the proof of this decomposition (in Section~\ref{sec:algebra}) from our algorithmic framework that uses it in a black-box manner (in Section~\ref{sec:algo}).
We devote Section~\ref{sec:ai} to discuss and reflect in detail on our methodology of AI-use.

\paragraph{Consequences for exact exponential-time algorithms.}
Our result also improves the running time of exact algorithms for several NP-hard optimization problems. Williams~\cite{Williams2005}, and subsequently Koivisto~\cite{Koivisto2006}, showed that Max-2-CSP on $n$ Boolean variables can be solved in $2^{\omega n/3+o(n)}$ time using fast matrix multiplication. Their framework includes Max-Cut and Max-2-SAT, as well as counting the number of optimal solutions. We improve this bound to
\[
    2^{(\omega/3-c)n+o(n)}
\]
for some absolute constant $c>0$, independent of $\omega$. Indeed, we partition the variables into nine groups, regard each assignment to a group as a vertex, and encode the constraints by unary and pairwise weights of a weighted $K_9$ instance. Our algorithm then gives the stated improvement, including when $\omega=2$. As in~\cite{Koivisto2006}, for polynomially bounded integer scores we can encode the objective values in the weights so that the resulting sum also determines the number of assignments attaining each score. Thus, the improvement applies to exact optimization and counting optimal solutions for general Max-2-CSP, including unrestricted dense instances.

The improvement also extends to several related counting problems. In particular, it gives the same exponent saving for counting induced copies of any fixed $k$-vertex graph, rather than only cliques~\cite{CurticapeanDoringNeuen2025}. Moreover, Scott and Sorkin~\cite{ScottSorkin2009} extended the matrix-multiplication approach to polynomial constraint satisfaction problems, including computing Ising partition functions, counting cuts of every size, and solving graph bisection. Our weighted counting algorithm improves the corresponding general bounds to $2^{(\omega/3-c)n+o(n)}$.

\section{Overview and organization}\label{sec:overview}

We give a brief overview of the algorithm, starting with the simpler case where $\omega>2$. Our algorithm improves the standard $n^{3\omega}$ running time for counting weighted copies of any \emph{pattern} graph on nine vertices. Rather than designing the algorithm specifically for $K_9$, we prove the improvement simultaneously for all nine-vertex pattern graphs, by induction on the number of edges. For sufficiently sparse patterns, the counting problem is easy. For denser patterns, we will reduce the problem to counting copies of patterns with fewer edges, together with one additional term that can be computed faster using square matrix multiplication as a black-box.

The standard $n^{3\omega}$ algorithm a-la \cite{NesetrilPoljak1985} partitions the nine vertices of the pattern graph $H$ into three sets of three. Via standard color-coding we can assume that the host graph $G$ is partitioned to three node parts $V_1,V_2,V_3$ where $V_i$ corresponds to the $i$th triple of vertices of $H$. Then to compute the weighted sum of the copies of $H$ in $G$, we create three $n^3\times n^3$ matrices $M_{12},M_{23},M_{31}$ encoding the weights of the triples inside each $V_i$ and where $M_{ij}$ encode the weights of the edges between every two triples in $V_i\times V_j$, and we finish by computing the trace of the matrix product $M_{12}M_{23}M_{31}$.

The original Ne\v{s}etril-Poljak style reduction leaves a lot of freedom how we pack the edge weights into the matrices $M_{ij}$. While it is clear that edges between two different parts $V_i$ and $V_j$ in $G$ must be represented in $M_{ij}$, there is freedom how the edge weights of a triple inside a particular $V_i$ are distributed between the two matrices $M_{i (i+1)}$ and $M_{i (i-1)}$ that see $V_i$.

This freedom is what our algorithm exploits. Suppose that $H$ contains two edges $xz$ and $yz$ that share the vertex $z$; we call such a $2$-path a \emph{wedge}. We choose the partition so that $x,z,y$ form one of the three triples, say the one corresponding to $V_1$, and we split the wedge between the two matrices that see $V_1$: the weight of $xz$ is placed in $M_{12}$ and the weight of $yz$ in $M_{31}$. We call these the first and the second matrix, respectively.

In the trace of $M_{12}M_{23}M_{31}$, the assignment to the triple of $V_1$ is both the row index of $M_{12}$ and the column index of $M_{31}$. The trace forces the two to coincide. We make this constraint explicit for $x$ and $y$. Let $s,u,v,t$ be vertices of the host graph $G$: $s$ and $u$ are the vertices to which the first matrix maps $x$ and $y$, respectively, and $v$ and $t$ are the vertices to which the second matrix maps $x$ and $y$, respectively. The middle vertex $z$ of the wedge is not split: both matrices see the same assignment to $z$, which, with a slight abuse of notation, we also denote by $z$. The standard algorithm requires the two matrices to map $x$ to the same vertex ($s=v$) and $y$ to the same vertex ($u=t$). We express this by inserting, between $M_{31}$ and $M_{12}$, an identity matrix $I$ whose rows are indexed by $(v,t)$ and whose columns are indexed by $(s,u)$; this leaves the trace unchanged.

\begin{figure}[t]
\centering
\small
\newcommand{\swaphl}[1]{\textcolor{red!75!black}{#1}}
\tikzset{
  swapvertex/.style={circle,draw,minimum size=6.5mm,inner sep=1pt},
  swapfirst/.style={thick,blue!70!black},
  swapsecond/.style={thick,dashed,orange!85!black},
  swaplab/.style={font=\scriptsize,inner sep=1.5pt}}
$\displaystyle
\underbrace{\bigl[\,w_{yz}(1,z)\ \ \swaphl{w_{yz}(2,z)}\ \ w_{yz}(1,z)\ \ w_{yz}(2,z)\,\bigr]}_{M_{31}}
\underbrace{\left[\begin{array}{@{}cccc@{}}
 1 & \cdot & \cdot & \cdot\\
 \cdot & \swaphl{\mathbf{1}} & \cdot & \cdot\\
 \cdot & \cdot & 1 & \cdot\\
 \cdot & \cdot & \cdot & 1
\end{array}\right]}_{I}
\underbrace{\left[\begin{array}{@{}c@{}} w_{xz}(1,z)\\ \swaphl{w_{xz}(1,z)}\\ w_{xz}(2,z)\\ w_{xz}(2,z)\end{array}\right]}_{M_{12}}
=\sum_{s,u}w_{xz}(s,z)\,w_{yz}(u,z)$

\bigskip

$\displaystyle
\underbrace{\bigl[\,w_{yz}(1,z)\ \ w_{yz}(2,z)\ \ \swaphl{w_{yz}(1,z)}\ \ w_{yz}(2,z)\,\bigr]}_{M_{31}}
\underbrace{\left[\begin{array}{@{}cccc@{}}
 1 & \cdot & \cdot & \cdot\\
 \cdot & \cdot & 1 & \cdot\\
 \cdot & \swaphl{\mathbf{1}} & \cdot & \cdot\\
 \cdot & \cdot & \cdot & 1
\end{array}\right]}_{P}
\underbrace{\left[\begin{array}{@{}c@{}} w_{xz}(1,z)\\ \swaphl{w_{xz}(1,z)}\\ w_{xz}(2,z)\\ w_{xz}(2,z)\end{array}\right]}_{M_{12}}
=\sum_{s,u}w_{xz}(s,z)\,w_{yz}(s,z)$

\bigskip

\begin{tabular}{c@{\hspace{3em}}c}
\begin{tikzpicture}[x=1cm,y=1cm,baseline={([yshift=-.5ex]current bounding box.center)}]
  \node[swapvertex] (s) at (0,0.8) {$s$};
  \node[swapvertex] (z) at (1.0,0) {$z$};
  \node[swapvertex] (u) at (2.0,0.8) {$u$};
  \draw[swapfirst] (s) -- node[swaplab,below left] {$w_{xz}(s,z)$} (z);
  \draw[swapsecond] (u) -- node[swaplab,below right] {$w_{yz}(u,z)$} (z);
\end{tikzpicture}
&
\begin{tikzpicture}[x=1cm,y=1cm,baseline={([yshift=-.5ex]current bounding box.center)}]
  \node[swapvertex] (s) at (0,0.8) {$s$};
  \node[swapvertex] (z) at (1.0,0) {$z$};
  \node[swapvertex] (u) at (2.5,0.8) {$u$};
  \draw[swapfirst] (s) to[bend left=25] node[swaplab,above right] {$w_{xz}(s,z)$} (z);
  \draw[swapsecond] (s) to[bend right=25] node[swaplab,below left] {$w_{yz}(s,z)$} (z);
  \draw[-{Latex},thick] (3.0,0.4) -- node[swaplab,above] {\emph{fold }} (3.7,0.4);
  \node[swapvertex] (s2) at (4.2,0.8) {$s$};
  \node[swapvertex] (z2) at (5.2,0) {$z$};
  \node[swapvertex] (u2) at (7.9,0.8) {$u$};
  \draw[very thick] (s2) -- node[swaplab,above right] {$w_{xz}(s,z)\,w_{yz}(s,z)$} (z2);
\end{tikzpicture}
\\[4pt]
{\footnotesize with $I$} & {\footnotesize with $P$}
\end{tabular}
\caption{The swap and the fold on a small example with host vertices $\{1,2\}$. We fix the assignments to $V_2$ and $V_3$ and the value of $z$, and keep only the factors coming from the two wedge edges. Since $yz$ is stored in $M_{31}$, the relevant row of $M_{31}$ is indexed by the pair $(v,t)$ to which $M_{31}$ maps $(x,y)$, and its entry is $w_{yz}(t,z)$; since $xz$ is stored in $M_{12}$, the relevant column of $M_{12}$ is indexed by the pair $(s,u)$ to which $M_{12}$ maps $(x,y)$, and its entry is $w_{xz}(s,z)$. Both are ordered $(1,1),(1,2),(2,1),(2,2)$, and dots are zeros. The identity $I$ pairs each $(s,u)$ with $(v,t)=(s,u)$, so the two edges depend on different host vertices and form the wedge $s$--$z$--$u$ (bottom left; $xz$ solid, $yz$ dashed). The swap $P$ pairs $(s,u)$ with $(v,t)=(u,s)$, so both edges depend on $(s,z)$ and fold into a single edge (bottom right). The highlighted entries follow the term $(s,u)=(1,2)$, which contributes $w_{xz}(1,z)\,w_{yz}(2,z)$ under $I$ and $w_{xz}(1,z)\,w_{yz}(1,z)$ under $P$.}
\label{fig:swap-fold-example}
\end{figure}
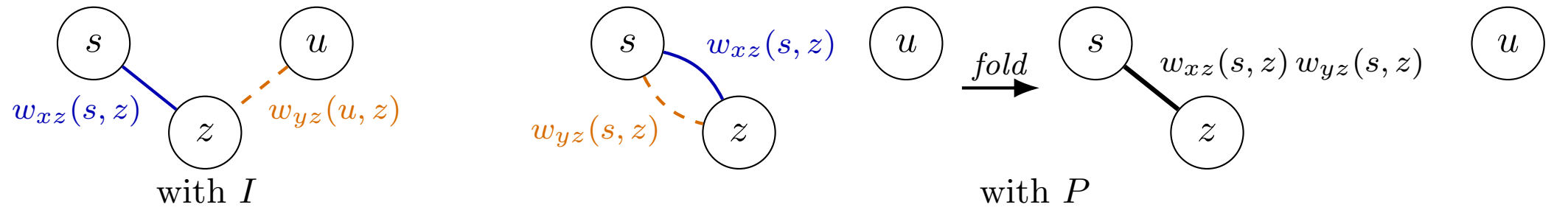

Instead, consider replacing this identity matrix with another permutation matrix, which we call the \emph{swap operator} $P$, that crosses these constraints: it imposes $s=t$ and $u=v$, so that the vertex used for $x$ by the first matrix is the one used for $y$ by the second, and vice versa. Since the edge $xz$ is in the first matrix and the edge $yz$ is in the second, replacing $I$ by $P$ transforms their contribution from $w_{xz}(s,z)w_{yz}(t,z)$ into
\[
 w_{xz}(s,z)w_{yz}(s,z).
\]
Thus the two edge weights can be replaced by a single weight on the edge~$xz$. All other weights still depend on at most two variables, so the resulting sum is a weighted sum problem for a pattern with at least one fewer edge. We refer to this operation as \emph{folding} the wedge. Figure~\ref{fig:swap-fold-example} illustrates the swap and the fold on a small example.

Of course, replacing $I$ by $P$ changes the answer. However, noting that $I=P+(I-P)$ allows us to compute the original sum by adding the contribution of the swap and the contribution of $I-P$. The first can be computed recursively on a sparser pattern. For the second, we can take advantage of the fact that $I-P$ has rank only $n(n-1)/2$, rather than $n^2$, which translates to lowering one of the three dimensions in the matrix multiplication instance we need to solve. Unfortunately, this is still $\Omega(n^2)$ and does not give a polynomial improvement in the running time.

We generalize the above attempt with a stronger version of this decomposition. We show that, for every constant $\varepsilon>0$, there exists a constant $\delta>0$ such that
\[
 I=R+\sum_{j=1}^{M}c_jS_j,
 \qquad
 M\le n^\varepsilon,
 \qquad
 \rank(R)\le n^{2-\delta}.
\]
Where the matrices $S_j$ are \emph{partial swaps}: a generalized version of the above swap that still allows folding an edge (and thus reducing the number of edges in the pattern) in a similar manner.

Let us see why this decomposition is sufficient when $\omega>2$. The contribution of $R$ can be computed by replacing one of the three matrix dimensions, originally $n^3$, by $n\cdot\rank(R)\le n^{3-\delta}$. By partitioning the other two dimensions into blocks and using square matrix multiplication, the resulting contraction can be evaluated in
\[
 n^{3\omega-(\omega-2)\delta+o(1)}
\]
time. When $\omega>2$, this gives a polynomial improvement over $n^{3\omega}$. We can also afford to explicitly construct the original matrices and apply the low-rank decomposition, since their total size is only $n^6<n^{3\omega}$.

The remaining contribution consists of at most $n^\varepsilon$ recursive calls to patterns with fewer edges. Suppose inductively that we have already obtained a polynomial saving $n^\alpha$ for all these patterns. We can choose $\varepsilon<\alpha$, so that even after making all the recursive calls we retain a positive saving. The algebraic decomposition guarantees that this choice still gives some $\delta>0$, however small $\varepsilon$ is. Since the number of edges in a nine-vertex pattern graph is at most $36$, the induction has only constantly many steps, and we obtain a polynomial improvement for every pattern. Thus, as long as $\omega>2$, the decomposition essentially gives the entire algorithm.

We state the decomposition formally in Section~\ref{sec:decomposition}, and give a short self-contained proof in Section~\ref{sec:algebra}. The algebraic construction itself is essentially a binary specialization of a matrix decomposition from the recent OpenAI result on the Exponential Time Hypothesis~\cite{OpenAIETH2026}, based on polynomial interpolation techniques previously used in matrix rigidity~\cite{AlmanWilliams2017}. We isolate the particular properties needed for our algorithm and prove them directly, using little more than repeated applications of $I=P+(I-P)$, elementary polynomial interpolation, and a concentration bound for the binomial distribution. 
We essentially attribute this decomposition to ideas appearing in the OpenAI manuscript~\cite{OpenAIETH2026}, and thus our main algorithmic contribution is in using this decomposition to cheaply reduce weighted graph counting to sparser patterns; This is our clique counting algorithm of Section~\ref{sec:algo}.

There is one additional complication: we want an improvement that holds independently of the value of $\omega$, even if $\omega=2$. The ``fold one wedge at a time to remove an edge'' argument sketched above does not achieve this, since reducing only one of the three matrix dimensions gives a saving proportional to $\omega-2$. Moreover, the explicit construction of the original matrices already costs $n^6$. To overcome both problems, we apply the decomposition simultaneously to two vertex-disjoint wedges, placed in two different sets of three vertices. The contribution in which both operators are replaced by their low-rank parts now has matrix dimensions $(n^{3-\delta},n^{3-\delta},n^3)$, and can be evaluated in $n^{3\omega-(\omega-1)\delta+o(1)}$ time, giving a saving even when $\omega=2$. The other terms are reduced to sparser patterns by applying partial swaps, as before. If the pattern does not contain two vertex-disjoint wedges, we show that it can already be counted in $O(n^5)$ time. We prove this stronger two-wedge reduction in Section~\ref{sec:two-wedges}, and complete the induction in Section~\ref{sec:algorithm}.

\section{Improved algorithm for counting weighted cliques}
\label{sec:algo}
Let $H$ be a \emph{pattern} graph on a fixed set of nine labeled vertices. For each vertex $i$ we are given a vertex-weight function $u_i:[n]\to\mathbb R$, and for each edge of the pattern graph $ij\in E(H)$ we are given an edge-weight function $w_{ij}:[n]^2\to\mathbb R$. We wish to compute
\begin{equation}\label{eq:graph-sum}
 Z_H=\sum_{a_1,\ldots,a_9\in[n]}
 \prod_{i=1}^9u_i(a_i)\prod_{ij\in E(H)}w_{ij}(a_i,a_j).
\end{equation}
All weights are arbitrary, and missing edges are interpreted as having weight one. 

The above is a standard and natural generalization of the k-Clique counting (and hence also detection) problem. 
If we set $H=K_9$, the adjacency matrix of a loopless graph as all the edge-weight functions~$w_{ij}$, and the constant one as all the vertex-weight function, then $Z_H/9!$ is the number of nine-cliques.
The generalization to all nine-vertex patterns~$H$ rather than only the clique~$K_9$ will be useful algorithmically, because we will iteratively construct improved algorithms to each of these patterns by making recursive calls to the same problem on sparser patterns.

The standard upper bound for detecting or counting 9-cliques works as-is also for computing \eqref{eq:graph-sum}: 
We partition the nine vertices of~$H$ into three sets $V_1,V_2,V_3$ of size three. Let $A,B,C\in[n]^3$ denote assignments to these sets. 
Every edge has both endpoints in at most two of the three sets, and its weight can therefore be assigned to one of three matrices $M_{12}(A,B)$, $M_{23}(B,C)$, $M_{31}(C,A)$. 
Unary weights are assigned to any one of the two matrices containing their corresponding vertex, with each weight appearing exactly once.
This gives
\begin{equation}\label{eq:triangle}
 Z_H=\sum_{A,B,C\in[n]^3}M_{12}(A,B)M_{23}(B,C)M_{31}(C,A)
 =\operatorname{tr}(M_{12}M_{23}M_{31}).
\end{equation}
The matrices have order $n^3$, so using fast matrix multiplication to compute the right-hand trace gives the $n^{3\omega}$ upper bound. Note that the explicit construction of the matrices costs $O(n^6)$ time, which is not larger than the first term as~$\omega\geq 2$.

\begin{figure}[t]
\centering
\begin{tikzpicture}[every node/.style={circle,draw,inner sep=1.7pt}, x=1cm,y=1cm, baseline=(current bounding box.center)]
  \node[draw=none] at (1.5,1.95) {$V_1$};
  \node[draw=none] at (5.0,1.95) {$V_2$};
  \node[draw=none] at (8.5,1.95) {$V_3$};
  \draw[rounded corners] (0,0) rectangle (3,1.7);
  \draw[rounded corners] (3.5,0) rectangle (6.5,1.7);
  \draw[rounded corners] (7,0) rectangle (10,1.7);

  \node (x1) at (0.6,1.1) {$x_1$};
  \node (z1) at (1.5,0.45) {$z_1$};
  \node (y1) at (2.4,1.1) {$y_1$};
  \draw[thick] (x1)--(z1)--(y1);

  \node (x2) at (4.1,1.1) {$x_2$};
  \node (z2) at (5.0,0.45) {$z_2$};
  \node (y2) at (5.9,1.1) {$y_2$};
  \draw[thick] (x2)--(z2)--(y2);

  \node (b1) at (7.6,1.1) {$a$};
  \node (b2) at (8.5,0.45) {$b$};
  \node (b3) at (9.4,1.1) {$c$};
\end{tikzpicture}
\caption{The standard reduction partitions the nine pattern vertices into three sets of three. When the pattern contains two vertex-disjoint wedges, we place one wedge in $V_1$ and one in $V_2$. The third set $V_3$ is arbitrary.}
\label{fig:partition}
\end{figure}
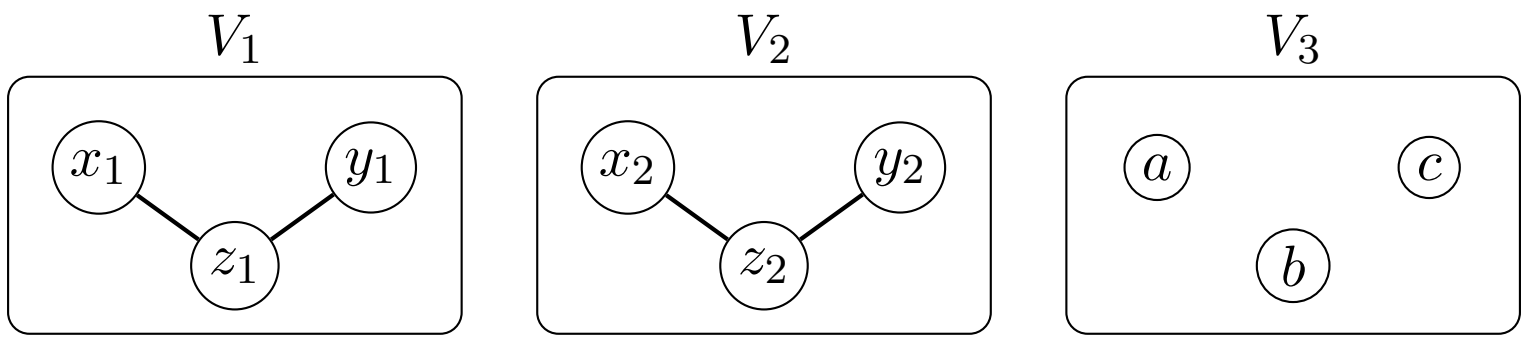

\subsection{Wedges and local swaps}\label{sec:prelim}

Our algorithmic framework is as follows: we will prove, inductively, that the above baseline of $n^{3\omega}$ can be polynomially improved for every pattern graph $H$; starting from the sparsest patterns as our base case and reducing solving the problem for a pattern~$H$ to solving it on patterns~$H'$ with fewer edges as the inductive step.

A \emph{wedge} is a triple of distinct vertices $x,z,y$ for which $xz,yz\in E(H)$. We do not require $xy$ to be absent (nor to exist). Two wedges are disjoint if they have no common vertex.

Our base case, that is, the set of patterns we can immediately give an improved algorithm for, is the family of all patterns $H$ that do not contain two vertex disjoint wedges. 

\begin{lemma}\label{lem:easy}
If $H$ contains no two vertex-disjoint wedges, its weighted counting sum can be computed in $O(n^5)\leq O(n^{3\omega -1})$ time. 
\end{lemma}
\begin{proof}
If $H$ contains no wedge, each connected component is an edge or an isolated vertex, so the sum factors into one- and two-variable sums. Otherwise fix any wedge $x,z,y$ and enumerate its $n^3$ assignments. The subgraph of $H$ induced by the remaining six vertices cannot contain another wedge, and is therefore a matching together with isolated vertices. After fixing $x,z,y$, all edges incident to those three vertices become unary weights on the remaining vertices. The remaining sum can be evaluated in $O(n^2)$ operations for each fixed assignment, giving the stated bound.
\end{proof}

We next describe a local operation on one wedge, which we call a \emph{swap}, and is meant to reduce the number of edges in the pattern by at least one. 
Place the wedge vertices $x,z,y$ in the same three-vertex set among $A,B,C$ in the reduction above. 
This set appears in two of the three matrices. 
In the first occurrence, use values $(s,u)$ to denote the assignment for the pair $(x,y)$; in the second, use values $(v,t)$ to denote it. 
We can, without changing anything, place the identity matrix  between these two matrices, as multiplying with it does not change the result.
We can then interpret this identity matrix as if it ``forces'' an equality between the assignments~$(s,u)$ and~$(v,t)$ to~$(x,y)$ in both matrices.
Indexing this matrix's rows and columns by these assignment, we write it as
\begin{equation}\label{eq:identity}
 I_{(s,u),(v,t)}=\one[s=v]\one[u=t].
\end{equation}
$I$ is simply the identity matrix of order $n^2$. 
Note that we are not planning to touch the third vertex (corresponding to~$z$) so we do not bother writing it down, but every time we write a matrix of dimension~$n^2$ we mean that the third variable~$z$ maintains equality (equivalently, the matrix should be tensorized with~$I_n$).

Consider what happens if we replace~$I$ with the following matrix $P$, which we call the \emph{swap operator},
\begin{equation}\label{eq:swap}
 P_{(s,u),(v,t)}=\one[s=t]\one[u=v].
\end{equation}
This exchanges the two values across the occurrences of the pair. It is a permutation matrix, and unlike~$I$, it \emph{does} change the computed value. 

Before we discuss the relation between the new quantity we compute after this replacement and what we actually want to compute, we discuss the implications of applying this swap operator on the pattern graph~$H$.

Intuitively, this swap \emph{folds} a wedge by swapping the two pivot values~$x,y$ across their occurrences in adjacent matrices. The two edge-weights $w_{xz}(s,z)$ and $w_{yz}(t,z)$ become a single edge-weight on $(s,z)$ after we impose that $s=t$.

\begin{figure}[!htbp]
\centering
\begin{tikzpicture}[
  x=1cm,y=1cm,
  vertex/.style={circle,draw,minimum size=7mm,inner sep=1pt},
  >=Latex
]
  \node[draw=none] at (1.4,1.85) {before};
  \node[vertex] (x) at (0.35,1.05) {$x$};
  \node[vertex] (z) at (1.4,0.35) {$z$};
  \node[vertex] (y) at (2.45,1.05) {$y$};
  \draw[thick] (x)--(z)--(y);

  \draw[->,thick] (3.25,0.85)--(4.45,0.85);
  \node[draw=none] at (3.85,1.23) {swap};

  \node[draw=none] at (6.3,1.85) {after};
  \node[vertex] (s) at (5.2,1.05) {$s$};
  \node[vertex] (zz) at (6.25,0.35) {$z$};
  \node[vertex] (u) at (7.3,1.05) {$u$};
  \draw[thick] (s)--(zz);
\end{tikzpicture}
%\caption{Folding a wedge by swapping the two pivot values across their occurrences in adjacent matrices. The two edge factors $w_{xz}(s,z)$ and $w_{yz}(t,z)$ become a single edge factor on $(s,z)$ after imposing $s=t$. Only the two wedge edges are shown; the second new pivot variable $u=v$ remains present, and all other edge factors remain unary or pairwise.}
%\label{fig:swap}
\end{figure}

To see it, place the edge $xz$ in the first matrix and $yz$ in the second. Their contribution is $w_{xz}(s,z)w_{yz}(t,z)$. Under \eqref{eq:identity} these are the original two edge weights. Under \eqref{eq:swap}, the constraint $s=t$ turns their product into
\begin{equation}\label{eq:fold}
 w_{xz}(s,z)w_{yz}(s,z),
\end{equation}
which can now be written as \emph{one} edge weight on the pair $(s,z)$. 
That is, we no longer need edge-weights on the second edge.
Note that the other new variable $u=v$ remains a separate vertex. All remaining original edge weights still depend on at most two variables. 
Thus the exchange produces a weighted graph sum in which the two edges of the wedge have become one. In particular, we reduced the number of edges in the pattern by one.
We will refer to this operation as \emph{folding} the wedge.

This observation already gives us a promising approach to attempt:
If we write the identity matrix as~$I=P+(I-P)$, we can reduce solving our original problem to computing two sums (and then adding them): the first sum, uses exactly our swap operator~$P$ instead of~$I$, and in turn corresponds to solving the problem on a sparser pattern - which we assume inductively already has a faster algorithm.
The other sum, uses the matrix~$(I-P)$ rather than~$I$ which also gives a slight improvement: for the~$n$ out of the~$n^2$ index pairs in which~$x=y$, the swap operator does nothing; in particular,~$I-P$ has~$n$ all-zero rows and thus its rank is $n(n-1)/2 < n^2$.
Replacing~$I$ with a low-rank matrix is also very useful algorithmically when computing the final matrix multiplications: it reduces one of the three dimensions in the matrix multiplication from~$n^3$ to~$n\cdot\text{rank}$.
Unfortunately, the rank of~$(I-P)$ is still~$\Omega(n^2)$, so this replacement doesn't result in any asymptotic saving.

On the other hand, this already motivates our entire algorithm:
Instead of decomposing~$I$ to merely one swap~$P$ and the remainder~$(I-P)$, we would decompose it to a combination of a larger number of swaps such that the overall remainder would be of sufficiently low rank.
Intuitively, if we know we can get a polynomial saving of~$n^{\alpha}$ on each of the sparser patterns, we can allow ourselves to solve reduced problems on polynomially many swap operators~$n^\varepsilon$ as long as~$\varepsilon<\alpha$.
We will show that a decomposition with any polynomial number~$n^\varepsilon$ of swap$^\star$ operators always exists that reduces the rank of the overall remainder multiplicatively by some \emph{polynomial} $n^\delta$.
The little asterisk is that the decomposition would not be to actual swap operators as we defined them above, but to a slightly more general version of them we will call partial swap, and is going to function in exactly the same manner.

\paragraph{Why do we ask for two disjoint wedges rather than one?}
It is worth explaining why we would simultaneously apply ``two iterations'' of our reduction to two disjoint wedges, rather than just doing so one by one. 
Suppose first that we apply it to a single wedge, placed in the first of the three blocks in~\eqref{eq:triangle}. In the low-rank contribution, the first block has at most $n^{3-\delta+o(1)}$ possible assignments, while the other two blocks still have $n^3$ assignments each. Thus, the resulting triangle contraction has dimensions
\[
    (n^{3-\delta},n^3,n^3).
\]
In particular, one of the input matrices is already of size~$n^3$-by-$n^3$ which has~$n^6$ entries.
While this is polynomially smaller than $n^{3\omega}$ when $\omega>2$, it prevents any savings when $\omega=2$. 

Applying the decomposition to two disjoint wedges avoids this problem. In the contribution where both operators are replaced by their low-rank parts, two of the three blocks have at most $n^{3-\delta+o(1)}$ assignments. 
Hence, we will be able to avoid paying~$n^6$ in the low-rank case, which provides an improvement even when $\omega=2$. 

\subsection{Partial swaps and a decomposition using them}
\label{sec:decomposition}

We first define the slightly more general swap operators that we will use. For simplicity, assume that $n=2^\ell$, so we can view every element of $[n]$ as a binary string of length $\ell$. This assumption is without loss of generality, since we can pad $n$ to the next power of two.

A \emph{partial swap} is defined by choosing a set $J\subseteq[\ell]$ of bit positions and fixing the values of the four bits $s_i,u_i,v_i,t_i$ for every $i\in J$. On all other positions, we impose the same constraints as in the full swap~\eqref{eq:swap}, namely $s_i=t_i$ and $u_i=v_i$. Formally, for fixed values $\rho_i\in\{0,1\}^4$ for each $i\in J$, the corresponding matrix is
\[
 S_{(s,u),(v,t)}
 =
 \prod_{i\notin J}\one[s_i=t_i]\one[u_i=v_i]
 \prod_{i\in J}\one[(s_i,u_i,v_i,t_i)=\rho_i].
\]
In particular, the original swap $P$ is the special case $J=\emptyset$.

As in the case of the full swap, we assign the two edges $xz$ and $yz$ to the two different matrices containing the wedge, and keep this assignment fixed throughout the decomposition. We next show that partial swaps retain the edge-folding property of the original swap.

\begin{lemma}\label{lem:folding}
Replacing the identity on the pivot pair of a wedge by any partial swap, with the edge weights assigned as above, gives a weighted graph-counting sum for a pattern with at least one fewer edge. The new weight functions can be constructed in $O(n^2)$ time.

Moreover, applying partial swaps simultaneously to two vertex-disjoint wedges reduces the number of edges by at least two.
\end{lemma}

\begin{proof}
Let $J$ be the set of fixed bit positions of the partial swap. Outside $J$, the constraints are $s_i=t_i$ and $u_i=v_i$, while all four bits are fixed on $J$. Thus, there are two free strings $p,q$, such that both $s,t$ are determined by $p$, and both $u,v$ are determined by $q$.

In particular, the two edges of the wedge contribute
\[
\widetilde w(p,z)=w_{xz}(s(p),z)w_{yz}(t(p),z),
\]
which is a single edge-weight function. Every other weight still depends on at most two variables, so the result is a weighted counting sum for a pattern with at least one fewer edge. The free strings may range over fewer than $n$ values, but can be regarded as variables over $[n]$ by extending the corresponding unary weights by zero.

For two vertex-disjoint wedges, we can assign their edge weights and apply the partial swaps independently, reducing the number of edges by at least two. Since there are constantly many unary and pairwise weight functions, the new tables can be constructed in $O(n^2)$ time.
\end{proof}

We now state the algebraic decomposition that we need. It can be viewed as a special case of a construction from the recent OpenAI result on the Exponential Time Hypothesis~\cite{OpenAIETH2026}. We give a simple self-contained proof in Section~\ref{sec:algebra}.

\begin{lemma}[Rank-or-swap decomposition]
\label{lem:decomposition}
For every constant $\varepsilon>0$, there exists a constant $\delta>0$ such that the identity matrix of order $n^2$ can be written as
\[
 I=R+\sum_{j=1}^{M}c_jS_j,
 \qquad
 M\leq n^\varepsilon,
 \qquad
 \operatorname{rank}(R)\leq n^{2-\delta},
\]
where each $S_j$ is a partial swap and $c_j\in\mathbb R$.

Moreover, the decomposition is explicit, and $R$ admits an explicit rank-one decomposition
\[
 R=\sum_{\lambda\in\Lambda}
 d_\lambda b_\lambda b_\lambda^\top,
 \qquad |\Lambda|\leq n^{2-\delta},
\]
where each vector $b_\lambda$ has at most $n$ nonzero entries, which can be efficiently enumerated. The vectors belong to an explicit basis whose corresponding basis transformation requires only $O(n^2\log n)$ time.
\end{lemma}

Together with Lemma~\ref{lem:folding}, Lemma~\ref{lem:decomposition} allows us to replace an identity matrix by either a low-rank matrix or a small number of recursive calls on sparser patterns. It remains to show that the contribution of the low-rank matrix can be evaluated sufficiently fast. For this, we will use not only its rank bound, but also the additional structure of its rank-one decomposition: it will allow us to construct the required matrix inputs sufficiently fast to not lose the saving even when $\omega=2$.

It is worth noting that the decomposition above already essentially gives an improved algorithm if $\omega>2$, folding only a single wedge at a time. 
Indeed, replacing the identity on one pivot pair by $R$ reduces one of the three matrix dimensions from $n^3$ to $n^{3-\delta}$, while the other two remain $n^3$. 
The resulting matrix multiplication can be performed in
\[
n^{3\omega-(\omega-2)\delta+o(1)}
\]
time, by partitioning the two larger dimensions into blocks of size $n^{3-\delta}$ and then using square matrix multiplication on each pair of blocks.
We can also afford to construct the original matrices and apply the basis transformation in $n^{6+o(1)}$ time, since $\omega>2$. 
The remaining terms give at most $n^\varepsilon$ recursive calls on patterns with fewer edges. Choosing $\varepsilon$ sufficiently small at each step of the induction therefore gives an exponent saving whenever $\omega>2$.
The two-wedge reduction is therefore needed to obtain a saving independent of $\omega$, including the case $\omega=2$. Since the formal analysis is essentially the same, we prove only this stronger version.

\subsection{Applying the decomposition to two wedges}
\label{sec:two-wedges}

We now describe the algorithm for a pattern $H$ containing two vertex-disjoint wedges. Place the wedges in $V_1$ and $V_2$, as in Figure~\ref{fig:partition}, and assign the two edges of each wedge to different matrices as before.

Apply Lemma~\ref{lem:decomposition} to the pivot pair of each wedge, with~$\varepsilon>0$ to be chosen later (which will come from the saving proved by induction to the sparser patterns). For $i=1,2$, write
\[
 I_i=R_i+T_i,
 \qquad
 T_i=\sum_{j=1}^{M_i}c_{i,j}S_{i,j},
\]
where $M_i\le n^\varepsilon$, and $R_i$ has rank at most $n^{2-\delta}$. As before, these operators act only on the two pivot variables, while the third variable in each block is left unchanged.

We could expand the product $I_1I_2$ into the four terms $R_1R_2$, $R_1T_2$, $T_1R_2$, and $T_1T_2$. However, the terms containing both a low-rank operator and a partial swap would not directly correspond to weighted graph-counting instances of the form~\eqref{eq:graph-sum}. Instead, we use the equivalent identity
\begin{equation}\label{eq:two-wedge-expansion}
 I_1I_2=R_1R_2+T_1I_2+I_1T_2-T_1T_2.
\end{equation}

Consider first the last three terms. By expanding each $T_i$ into its partial swaps, we obtain at most
\[
 M_1+M_2+M_1M_2=O(n^{2\varepsilon})
\]
weighted graph-counting instances. Indeed, the term $T_1I_2$ applies a partial swap only to the first wedge, the term $I_1T_2$ applies one only to the second wedge, and the term $T_1T_2$ applies partial swaps to both. By Lemma~\ref{lem:folding}, every resulting instance has strictly fewer edges than $H$. The new weight functions can be constructed in $O(n^2)$ time per instance.

Thus, we have reduced the computation of $Z_H$ to $O(n^{2\varepsilon})$ recursive calls on sparser patterns, and one additional term corresponding to $R_1R_2$. We next show that this remaining term can be evaluated faster than the original matrix multiplication.

\begin{lemma}\label{lem:low-rank}
Assuming any matrix multiplication constant $\omega\geq 2$, the contribution of $R_1R_2$ in~\eqref{eq:two-wedge-expansion} can be computed in
\[
 n^{3\omega-(\omega-1)\delta+o(1)}
\]
time, including the construction of all the required matrices.
\end{lemma}

\begin{proof}
Recall that each low-rank operator admits a decomposition
\[
 R_i=\sum_{\lambda\in\Lambda}
 d_\lambda b_\lambda b_\lambda^\top,
 \qquad |\Lambda|\le n^{2-\delta}.
\]
Inserting these decompositions into~\eqref{eq:triangle} replaces each of the two pivot pairs by an index $\lambda\in\Lambda$. The third vertex of each wedge remains unchanged. Consequently, the first two blocks now have at most
\[
 D=n|\Lambda|\le n^{3-\delta}
\]
possible assignments each, while the third block still has $n^3$ assignments. The coefficients $d_\lambda$ can be absorbed into the entries of the resulting matrices.

If the matrices corresponding to this contraction were already given, we could evaluate it efficiently using square matrix multiplication. Indeed, the three dimensions are now $(D,D,n^3)$. Padding the first two dimensions to $n^{3-\delta}$, partitioning the third dimension into $O(n^\delta)$ blocks of the same size, and using square matrix multiplication on each block gives a running time of
\begin{equation}\label{eq:low-rank-time}
 n^\delta\cdot n^{\omega(3-\delta)+o(1)}
 =n^{3\omega-(\omega-1)\delta+o(1)}.
\end{equation}
In particular, this gives a universal polynomial saving of at least~$\delta$ in the exponent, which is independent of $\omega$.

However, we must also construct the matrices within this running time. As long as we want to be able to beat~$n^6$, we cannot simply construct the original matrices of order $n^3$ and apply the low-rank projections afterward, since this would already require $n^6$ time.

Consider a matrix containing one of the two blocks to which we apply the low-rank decomposition. Denote the assignments to this block by $(s,u,z)$, where $(s,u)$ is the pivot pair, and let $B\in[n]^3$ be the assignment to the other block. For a fixed $\lambda\in\Lambda$, the corresponding projected matrix has entries
\begin{equation}\label{eq:projected-matrix}
 \widetilde M((\lambda,z),B)
 =
 \sum_{s,u\in[n]}b_\lambda(s,u)M((s,u,z),B).
\end{equation}
Here we identify $b_\lambda$ with a function on $[n]^2$.

By Lemma~\ref{lem:decomposition}, $b_\lambda$ has at most $n$ nonzero entries. Enumerate these entries as
\[
 (s_j,u_j),\qquad j=1,\ldots,r_\lambda,
 \qquad r_\lambda\le n.
\]
We can therefore rewrite~\eqref{eq:projected-matrix} as
\[
 \widetilde M((\lambda,z),B)
 =
 \sum_{j=1}^{r_\lambda}
 b_\lambda(s_j,u_j)M((s_j,u_j,z),B).
\]

Recall that every entry of $M$ is a product of unary and pairwise weight functions. After substituting $s=s_j$ and $u=u_j$, each factor depends either on $(z,j)$, on $(j,B)$, or on $(z,B)$. We may therefore group the factors and write
\begin{equation}\label{eq:projected-factorization}
 \widetilde M((\lambda,z),B)
 =
 K_\lambda(z,B)
 \sum_{j=1}^{r_\lambda}
 A_\lambda(z,j)B_\lambda(j,B),
\end{equation}
where the coefficients $b_\lambda(s_j,u_j)$ are included in $A_\lambda$.

The sum on the right-hand side is a matrix multiplication between an $n\times r_\lambda$ matrix and an $r_\lambda\times n^3$ matrix. Since $r_\lambda\le n$, we can pad the first two dimensions to $n$, partition the second matrix into $n^2$ blocks of $n$ columns each, and use square matrix multiplication. This takes $O(n^{\omega+2})$ time for each $\lambda$. Constructing the matrices $A_\lambda,B_\lambda,K_\lambda$ and multiplying the result by $K_\lambda$ takes only $O(n^4)$ additional operations, which is covered by the same bound.

As there are at most $n^{2-\delta}$ choices of $\lambda$, we can construct the entire projected matrix in
\begin{equation}\label{eq:projection-cost}
 n^{\omega+4-\delta+o(1)}
\end{equation}
time. The same argument applies when the projected block occurs in the columns rather than the rows.

This handles the two matrices connecting a projected block to the unprojected third block. The remaining matrix, $M_{12}$, has both endpoints projected. To construct it, we first apply the above procedure to the pivot pair in $V_1$, without yet projecting the pivot pair in $V_2$. This produces a matrix of size at most
\[
 n^{3-\delta}\times n^3,
\]
containing at most $n^{6-\delta}$ entries.

We then apply the full basis transformation from Lemma~\ref{lem:decomposition} to the pivot pair in $V_2$, and retain only the indices in $\Lambda$. Since the transformation requires $O(n^2\log n)$ operations on each array of $n^2$ entries, the total cost is
\[
 n^{6-\delta+o(1)}.
\]
The two projections act on different variables, so this gives exactly the matrix obtained by applying both low-rank decompositions. In particular, we never need to construct the original $n^3$-by-$n^3$ matrix.

We have thus constructed all three matrices in $n^{\omega+4-\delta+o(1)}$ time, which is no larger than the multiplication time in~\eqref{eq:low-rank-time}.
\end{proof}

\subsection{Completing the induction}
\label{sec:algorithm}

We now combine the two parts to obtain the improved algorithm for every nine-vertex pattern.

Let $\mathcal T_m(n)$ denote the maximum running time of our algorithm over all weighted patterns on nine vertices with at most $m$ edges. We prove by induction on $m$ that there exists a constant $\alpha_m>0$, independent of $\omega$, such that
\begin{equation}\label{eq:induction-bound}
 \mathcal T_m(n)\le n^{3\omega-\alpha_m+o(1)}.
\end{equation}

The base case $m=0$ is immediate, and we may take $\alpha_0=1$. Suppose that the claim holds for $m-1$, with some $\alpha_{m-1}>0$. Choose a sufficiently small constant $\varepsilon>0$ such that
\[
 2\varepsilon<\frac{\alpha_{m-1}}2,
\]
and let $\delta>0$ be the corresponding constant from Lemma~\ref{lem:decomposition}.

Consider a pattern $H$ with at most $m$ edges. If it does not contain two vertex-disjoint wedges, Lemma~\ref{lem:easy} gives an $O(n^5)$ algorithm, which already satisfies~\eqref{eq:induction-bound}.

Otherwise, we apply the decomposition to two vertex-disjoint wedges as described above. By Lemma~\ref{lem:low-rank}, the contribution of $R_1R_2$ can be evaluated in
\[
 n^{3\omega-(\omega-1)\delta+o(1)}
\]
time. All the remaining terms reduce to $O(n^{2\varepsilon})$ weighted counting instances for patterns with fewer edges, each of which can be constructed in $O(n^2)$ time. We therefore obtain
\begin{equation}\label{eq:recursive-time}
 \mathcal T_m(n)
 \le
 n^{3\omega-(\omega-1)\delta+o(1)}
 +n^{2\varepsilon+o(1)}
 \bigl(\mathcal T_{m-1}(n)+O(n^2)\bigr).
\end{equation}

By the induction hypothesis and our choice of $\varepsilon$, the second term is bounded by
\[
 n^{3\omega-\alpha_{m-1}+2\varepsilon+o(1)}
 \le n^{3\omega-\alpha_{m-1}/2+o(1)}.
\]
For the first term, recall that $\omega\ge2$, so $\omega-1\ge1$. Its running time is therefore at most
\[
 n^{3\omega-\delta+o(1)}.
\]
Consequently, both terms have a polynomial saving over $n^{3\omega}$. For example, we may take
\[
 \alpha_m=\frac14\min\{\alpha_{m-1},\delta\}>0.
\]
This completes the induction.

Since a nine-vertex pattern graph has at most $\binom92=36$ edges, we obtain some constant $\alpha_{36}>0$ such that every weighted nine-vertex pattern can be counted in
\[
 n^{3\omega-\alpha_{36}+o(1)}
\]
time. The constant $\alpha_{36}$ depends only on the decomposition lemma and the number of induction steps, and in particular is independent of $\omega$.

The same result also gives an improvement for counting $k$-cliques whenever $k$ is divisible by nine. Indeed, write $k=9t$ and partition the $k$ vertices into nine sets of $t$ vertices each. We can regard an assignment to each set as a single value in a domain of size $N=n^t$, absorbing the edges within a set into unary weights and the edges between two sets into pairwise weights. Applying the algorithm above to this weighted nine-vertex instance gives
\[
 N^{3\omega-\alpha_{36}+o(1)}
 =n^{(\omega/3-\alpha_{36}/9)k+o(1)}.
\]
Thus, we obtain a saving proportional to $k$ over the standard $n^{\omega k/3}$ bound.

We remark that as the recursion depth is constant, all edge weights remain within polynomial bound of the original input ones, and thus arithmetic operations and word RAM are equivalent.

\section{Proof of the rank-or-swap decomposition}
\label{sec:algebra}

We now prove Lemma~\ref{lem:decomposition}. The construction is a binary specialization of the matrix decomposition used in the recent OpenAI result on the Exponential Time Hypothesis~\cite{OpenAIETH2026}, which builds on earlier polynomial interpolation techniques from matrix rigidity~\cite{AlmanWilliams2017}. 
A related decomposition using partial transposes, which is also a synthesis of the OpenAI paper, appears in the recent work of Alman et al.~\cite{AlmanLiPantPrattWilliams2026}. 
We give a direct proof starting from the elementary identity $I=P+(I-P)$.

\begin{proof}[Proof of Lemma~\ref{lem:decomposition}]
Let $n=2^\ell$, and identify the pair space $[n]^2$ with $\ell$ copies of the one-bit pair space $\{00,01,10,11\}$. On this four-dimensional space, the swap operator fixes $00,11$ and exchanges $01,10$. The four vectors
\[
 e_{00},\qquad e_{11},\qquad
 e_{01}+e_{10},\qquad e_{01}-e_{10}
\]
form an orthogonal eigenbasis. The first three have eigenvalue $1$ under the swap, and the last has eigenvalue $-1$.

For each position $i\in[\ell]$, let $P_i$ swap the $i$-th pair of bits, leaving all other positions unchanged, and put $C_i=I-P_i$. For $0\le m\le\ell$, let $W_m$ be the sum of all products that choose $P_i$ at exactly $m$ positions and $C_i$ at the others. Since $P_i+C_i=I$, we have
\begin{equation}\label{eq:words}
 \sum_{m=0}^{\ell}W_m
 =\prod_{i=1}^{\ell}(P_i+C_i)=I.
\end{equation}

The operators $W_m$ are diagonal in the tensor product of the one-bit eigenbases described above. A basis vector is specified, among other information, by the number $j$ of its antisymmetric factors $e_{01}-e_{10}$. Consider the generating polynomial
\[
 \sum_{m=0}^{\ell}z^mW_m
 =\prod_{i=1}^{\ell}(zP_i+C_i).
\]
On a basis vector with $j$ antisymmetric factors, its eigenvalue is
\[
 z^{\ell-j}(2-z)^j.
\]
Indeed, a symmetric factor contributes $z$, while an antisymmetric factor contributes $2-z$. In particular, the eigenvalue of $W_{\ell-h}$ is
\[
 (-1)^j(-2)^h\binom{j}{h}.
\]

Fix a sufficiently small constant $\eta>0$, to be chosen later, and let $d=\lfloor\eta\ell\rfloor$. We will write
\[
 I=R+T,\qquad
 T=\sum_{h=0}^{d}c_{\ell-h}W_{\ell-h},
 \qquad R=I-T.
\]
By the calculation above, the eigenvalue of $T$ on a type-$j$ basis vector is $(-1)^jQ(j)$, where
\[
 Q(j)=\sum_{h=0}^{d}c_{\ell-h}(-2)^h\binom{j}{h}.
\]
Since the polynomials $\binom{j}{h}$, for $0\le h\le d$, form a basis for polynomials of degree at most $d$, we can choose $Q$ to be any such polynomial.

Choose an integer $a$ such that the interval $[a,a+d]$ is centered around $\ell/4$, up to rounding, and set
\begin{equation}\label{eq:interpolate}
 Q(j)=(-1)^a\sum_{h=0}^{d}(-2)^h
 \binom{j-a}{h},
\end{equation}
where the binomial coefficients are interpreted as polynomials in $j$. For every integer $j\in[a,a+d]$, the sum ends at $h=j-a$, and the binomial theorem gives $Q(j)=(-1)^j$. Consequently, $T$ has eigenvalue one on all these basis vectors, and $R$ vanishes on them. The coefficients $c_m$ are obtained by expressing~\eqref{eq:interpolate} in the basis $\{\binom{j}{h}\}_{h=0}^d$, and can be computed explicitly in time polynomial in $\ell$.

\paragraph{Rank of $R$.}
There are exactly $\binom{\ell}{j}3^{\ell-j}$ product-basis vectors with $j$ antisymmetric positions. Their total number is $4^\ell=n^2$, and their normalized multiplicities are the probabilities of a random variable $J\sim\operatorname{Bin}(\ell,1/4)$.

Since $R$ vanishes on all basis vectors with $j\in[a,a+d]$, its rank is bounded by the number of vectors outside this interval. The interval is centered at $\ell/4$ and has width $\eta\ell-O(1)$, so every integer outside it is at distance at least $\eta\ell/2-O(1)$ from the mean. By the standard Hoeffding bound,
\[
 \Pr[|J-\ell/4|\ge t]\le 2e^{-2t^2/\ell}.
\]
Therefore,
\[
 \operatorname{rank}(R)
 \le n^2\exp(-\eta^2\ell/2+O(1))
 \le n^{2-\eta^2/4+o(1)}.
\]

\paragraph{Expansion into partial swaps.}
Every word in $T$ contains at most $d$ factors $C_i=I-P_i$. In the standard basis on a single bit pair, $C_i$ has exactly four nonzero entries. Expanding a word by choosing one of these entries at every position where $C_i$ appears gives a signed combination of operators that swap the two bit values at all other positions and fix all four bits at these positions. These are precisely the partial swaps defined in Section~\ref{sec:decomposition}.

The total number of partial swaps, counting multiplicity, is at most
\[
 \sum_{h=0}^{d}\binom{\ell}{h}4^h
 \le n^{H_2(\eta)+2\eta+o(1)},
\]
where $H_2$ denotes binary entropy. Since $H_2(\eta)+2\eta$ tends to zero as $\eta\to0$, for every constant $\varepsilon>0$ we can choose $\eta>0$ sufficiently small that the number of partial swaps is at most $n^\varepsilon$. The rank bound above then gives $\operatorname{rank}(R)\le n^{2-\delta}$ for some constant $\delta>0$.

\paragraph{Structure of the basis.}
Each one-bit eigenvector has at most two nonzero entries, so each product-basis vector has at most $2^\ell=n$ nonzero entries, which can be enumerated directly from its one-bit factors.

Since $R$ is diagonal in this orthogonal basis, we obtain the rank-one decomposition
\[
 R=\sum_{\lambda\in\Lambda}
 d_\lambda b_\lambda b_\lambda^\top,
 \qquad |\Lambda|=\operatorname{rank}(R)
 \le n^{2-\delta},
\]
where the norms of the basis vectors are absorbed into the coefficients $d_\lambda$.

Finally, the change of basis can be performed by applying the same four-dimensional transformation separately at each of the $\ell$ bit positions. Each position requires $O(n^2)$ arithmetic operations, so the entire transformation takes $O(n^2\log n)$ time, as required.
\end{proof}

\section{Discussion and possible improvements}
\label{sec:discuss}
We disproved the $k$-Clique conjecture by obtaining a universal constant improvement to the conjectured~$\frac{\omega}{3}$ coefficient in the running time exponent, which is independent of either~$k$ or the matrix multiplication exponent~$\omega\geq 2$.
Our guiding principle throughout this work and its writing was to obtain and present the preceding statement in the simplest and easiest to explain manner.
In particular, we did not make any attempt at optimizing the resulting constant---quite the contrary, we made it significantly worse for the sake of making the arguments cleaner.
For example, to be able to get a fixed improvement irrespective of the eventual~$\omega$, we never use rectangular matrix multiplication (but instead reduce it to square products); or, in the inductive argument of Section~\ref{sec:algo}, we analyze a universal bound on all patterns with the same number of edges rather than analyzing each pattern separately to optimize bounds (there is only a constant number of patterns, but an absurdly large one). 
Therefore, we do not state or claim any specific numerical bounds (which, with the presented framework, will seemingly range between minuscule without optimization to tiny with significant optimizations). 

We leave as interesting open problems obtaining a \emph{significantly} smaller exponent than~$\frac{\omega}{3}k$, and in particular the question of whether linear dependence on~$k$ in the exponent is even necessary. As the recent refutation of the Exponential Time Hypothesis \cite{OpenAIETH2026,ClaudeETH2026} suggests, large improvements in the exponent are possible.

\section{Reflection on Human-AI interaction}
\label{sec:ai}
This work was achieved by a combination of a traditional research process with extensive repeated use of a publicly available AI tool (OpenAI's ChatGPT-6, on the web interface without special harnesses, tools, token use, etc.).
We note that both of us have our memory feature turned on, so any result we attribute to the AI could have also been affected by previous projects, but nonetheless we will attempt to use this section to describe our AI-use methodology to the best of our understanding.

Our starting point was providing the AI model with previous ideas of ours for an approach to improve the running time for finding $5$-cliques specifically. We also gave it the context of recent breakthroughs in fine-grained complexity and in particular the ETH refutation of OpenAI~\cite{OpenAIETH2026}.
After testing some ideas in a back-and-forth conversation, it was eventually able to come up with the first improvement for $5$-cliques: it gave an improvement over the $\omega(2,1,2)$ exponent. This improvement
initially appeared rather complicated. While the model gave a generalization for some larger clique sizes, the improvements were miniscule and it initially did not report an improvement for clique sizes divisible by three, which is needed for the generalization to infinitely many values of~$k$ (since we want an improvement in terms of $\omega$ and not rectangular exponents).

%the model was not immediately able to generalize it to larger clique sizes (and in particular not to ).

What followed was a process of AI-assisted distillation and generalization of that~$K_5$ idea, which eventually resulted in our understanding of the algebraic decomposition of Section~\ref{sec:algebra}, which we could separate and cleanly define, as well as trace it to ideas from the OpenAI paper~\cite{OpenAIETH2026}. Similar ideas (plausibly originating from the same place) also appeared in the three recent AI-assisted refutations of the $k$-HyperClique conjecture~\cite{AlmanLiPantPrattWilliams2026,Fischer2026Hyperclique,ChanItakuraMaYu2026}.

Much of the eventual algorithmic framework that uses this algebraic decomposition came from us, and understanding the algebraic decomposition and how to use it allowed us to obtain an algorithm that generalizes to all large enough~$k$, gives an improvement despite only using square matrix multiplication as a black-box (rather than using rectangular matrix multiplication or attempting to directly `invent' new matrix-multiplication-like tensors; which would make a universal bound independent of~$\omega$ very unlikely), and is rather natural and simple when taking the algebraic decomposition as a given black-box.
We then also used iterative AI use to simplify the algebraic decomposition and write it in a purely combinatorial manner.

One of our (subjective) observations from this process, which we state as it might be useful for other researchers developing an AI-use methodology, is the following:
While we all know by now that the obfuscated and cumbersome proofs generated by AI are very difficult for us humans to read and comprehend, we were impressed that they are also difficult for the AI itself to understand. The models were consistently unable to develop further ideas generated by them, but when presented with a clean simplification of the ideas, they (as well as us) were suddenly much more able to take further steps.

Roughly, then, we view the algebraic decomposition of Section~\ref{sec:algebra} as almost-purely the AI's contribution (originating from~\cite{OpenAIETH2026}), and the algorithmic framework that uses it as a black-box in Section~\ref{sec:algo} as mainly ours.

\bibliographystyle{alphaurl} 
\bibliography{refs}
%\begin{thebibliography}{9}

%\end{thebibliography}
\end{document}